\documentclass[aip,cha,preprint]{revtex4-1}

\usepackage{graphicx}
\usepackage{amssymb}
\usepackage{color}
\usepackage{dcolumn}
\usepackage{bm}
\usepackage[final]{changes}

\newcommand{\figsizeone}{0.8}
\newcommand{\figsizetwo}{0.8}
\newcommand{\figsizethree}{0.6}

\newcommand{\figsizefive}{0.9}

\makeatletter
\let\Changes@Markup@Deleted\@gobble
\makeatother

\begin{document}

\draft
\title{Statistical properties of chaotic microcavities in small and large opening cases}
\author{Jung-Wan Ryu}
\address{Center for Theoretical Physics of Complex Systems, Institute for Basic Science (IBS), Daejeon 34051, Republic of Korea}
\author{Sang Wook Kim}
\email{swkim0412@pusan.ac.kr}
\address{Department of Physics, Kyung Hee University, Seoul 02447, Republic of Korea}

\begin{abstract}
We study the crossover behavior of statistical properties of eigenvalues in a chaotic microcavity with different refractive indices.
The level spacing distributions change from Wigner to Poisson distributions as the refractive index of a microcavity decreases.
We propose a non-hermitian matrix model with random elements describing the spectral properties of the chaotic microcavity, which exhibits the crossover behaviors as the opening strength increases.
\end{abstract}

\maketitle
\narrowtext

\begin{quotation}
Dielectric microcavities serve as useful platforms for studying quantum chaos in the case of large opening. However, there are few studies on statistical properties of eigenfunctions in dielectric microcavities. We study the distributions of the level spacings of the real parts and the probability distributions of the imaginary parts of complex eigenvalues for the stadium microcavities with different refractive indices and discuss the differences between the statistical properties of eigenfunctions in the microcavities with large and small refractive indices. We also propose the non-Hermitian matrix model corresponding to the chaotic microcavities.
\end{quotation}

\section{Introduction}

Every real quantum system is inevitably open since no information can be extracted from completely closed systems.
However, open quantum systems are very different from closed ones.
In the mathematical viewpoint, the closed quantum systems are described by usual Hermitian formalism,
while the open ones by non-Hermitian formalism \cite{Rot09,Moi11}.
In contrast to Hermitian Hamiltonian, a general non-Hermitian Hamiltonian has complex eigenvalues and
non-orthogonal eigenfunctions \cite{Ste72,Per96}.
The imaginary parts of the eigenvalues represent the decay rate of the corresponding eigenmode.
\added{In particular, the eigenfunction non-orthogonality
is responsible for high sensitivity of the decay rates to perturbations which has been
demonstrated both theoretically \cite{Fyo12} and experimentally \cite{Gro14}.}
One of the remarkable feature of non-Hermitian Hamiltonian is the existence of an exceptional point,
at which two complex eigenvalues coalesce, so do the corresponding eigenstates \cite{Kat66,Hei99,Hei00,Hei12}.
Recently, non-Hermitian Hamiltonian draws much attention due to its various applications,
e.g., optical systems with complex refractive indices \cite{El07,Mus08,Mak08}
and parity-time symmetric quantum Hamiltonian systems \cite{Ben98,Ben07}.

Statistical properties of spectra of closed quantum systems are important for studying quantum chaos
which describes the quantum mechanical behavior of classically chaotic systems \cite{Sto99}.
It has been found that the so-called random matrix theory (RMT)
which was originally introduced to model the nuclei of heavy atoms
successfully explains the statistical nature of the spectra of fully-chaotic systems \cite{Boh84}.
According to the RMT, the nearest neighbor spacing of eigenenergies exhibits Wigner distribution in chaotic systems,
while Poisson distribution in integrable systems.
The RMT predicts the eigenstates of fully chaotic systems are delocalized over the entire accessible energy surface in phase space
with some fluctuation described by Porter-Thomas distribution and locally look like random superpositions of the plane waves in coordinate space \cite{Ber77,Vor79}.
Different from most of such chaotic states, a few eigenstates appear to be localized
around unstable periodic orbits in the semiclassical regime, which is called as scarring \cite{Hel84,Bog88,Hel91}.

Dielectric microcavities have been extensively studied due to its wide range of applications \cite{Cha96,Vah03}.
\added{In particular, the statistics of complex eigenvalues for such systems is experimentally accessible \cite{Wan16}}.
Besides many useful applications, it provides the paradigm to study quantum chaos in open systems \cite{Cao15}, along with microwave billiards \cite{Sto90,Haa91}, quantum corrals \cite{Hel94}, \added{and quantum graphs or microwave networks \cite{Kot00,Sir10,All14}.}
Especially, focus lies at the ray-wave correspondence of eigenmodes in chaotic microcavities, where key issues are the frequent
occurrence of localized eigenmodes and the so-called universal directionality of far field emission \cite{Sch04,Lee05,Lee07}.
The eigenenergies of the chaotic microcavity are complex because the microcavity is intrinsically an open system.
It is reported that the distribution of the imaginary parts of complex eigenvalues of the eigenmodes with rather higher Q-factor,
defined as the real part versus the imaginary part, in chaotic microcavities is described by the fractal Weyl's law \cite{Wie08,Sch09}.
In the deformed rough microcavities, the distribution of the imaginary parts is strongly affected by dynamical localization \cite{Sta00}.
\deleted{Besides studies on eigenenergies in dielectric cavities, there have been many works on statistical properties of eigenvalues in open systems\added{\cite{Leh95, Fyo16}}.} Furthermore,
the Wigner surmise for open chaotic systems has also been derived analytically based upon two-level model with one-channel case,
which is generalized to N-channels with one free parameter of openness and tested experimentally \cite{Pol12}.
\replaced{T}{The spectral properties of microwave graphs \cite{All14} and t}he distribution of avoided level crossings \cite{Pol09} have also been studied in similar context.

Although the localized eigenstates such as scars do not quite often occur in the {\em closed} quantum systems with classically chaotic dynamics such as chaotic billiards,
they are more likely to be observed both experimentally and numerically in the dielectric microcavity systems \cite{Gma02,Lee02,Har03,Fan07a}.
This means that open systems such as dielectric mircocavities can show more frequently localized eigenstates than scars of closed systems.
In addition, in the recent experiments of deformed dielectric microcavities made of the material with a rather lower $n$, where $n$ is the index of refraction,
namely $n < 2$, not only the signatures of localized eigenstates are observed but the wide range of spectra also exhibit even equidistant spacing,
the characteristics of strongly localized modes on periodic orbits \cite{Leb06,Fan07b,Lee09,Dje09,Bog11,Bit12}.
These results imply that dielectric microcavities with small refractive indices show more localized eigenstates than those with large refractive indices (see Appendix A).
As a result, the dielectric microcavities with small refractive indices are qualitatively different from those with large refractive indices.

In this work,
we study the qualitative difference between statistical properties of complex eigenvalues in dielectric microcavities with large and small refractive indices,
meaning small and large opening cases, respectively.
We show that the statistical properties of complex eigenvalues of chaotic dielectric microcavities drastically changes as the opening becomes larger;
from the Wigner to Poisson distribution in the nearest neighbor spacings of the real parts.
We also investigate a plausible matrix model exhibiting the similar structure of the opening that the chaotic dielectric microcavities inherently pose.
Eigenfunctions of the matrix model exhibit corresponding crossover behaviors as the opening strength increases.

This paper is organized as follows.
In Sec. II, we show that the distributions of the level spacings of the real parts and the probability distributions of the imaginary parts
of eigenvalues for the stadium microcavities depend on the refractive indices.
In Sec. III, the matrix model corresponding to the chaotic microcavities with different refractive indices is introduced.
Finally, we summarize our results in Sec. IV.

\section{Microcavity}

Let us consider the statistical properties of complex eigenvalues of the stadium microcavities.
The resonances and the corresponding quasibound modes of a microcavity are obtained from solving the Helmholtz equation,
$[\nabla^2+n^2(\bold{r})k^2]\psi=0$, where $n$ is the refractive index of the dielectric exploited, by using the boundary element method \cite{Wie03}.
We focus on TM polarization where both the wavefunction $\psi$ and its normal derivative $\partial_{\nu}\psi$ are continuous across the boundary.
The $\psi$ corresponds to the $z$ component of the electric field $E_z$
when the cylindrical geometry of the cavity is assumed so that it is enough to take the cross sectional area
in $xy$-plane, into account to describe the $\psi$ \cite{Jac62}.
Here we consider the dielectric cavity whose boundary is given as Bunimovich stadium consisting of a square and two semicircles with radius $R$.
This is a paradigm model of quantum chaos \cite{Bun79}. Here we set $R=1$.
Due to the reflection symmetries with respect to both $x$ and $y$ axes, we consider the modes with only even-even parity without loss of generality.
Real and imaginary parts of complex $k$ represent frequency (energy) and decay (inverse lifetime) of the resonance, respectively.
In this section, we explore the eigenvalues in stadium microcavities with different refractive indices
which determine the openness of the microcavity.

\begin{figure}
\begin{center}
\includegraphics[width=\figsizeone\textwidth]{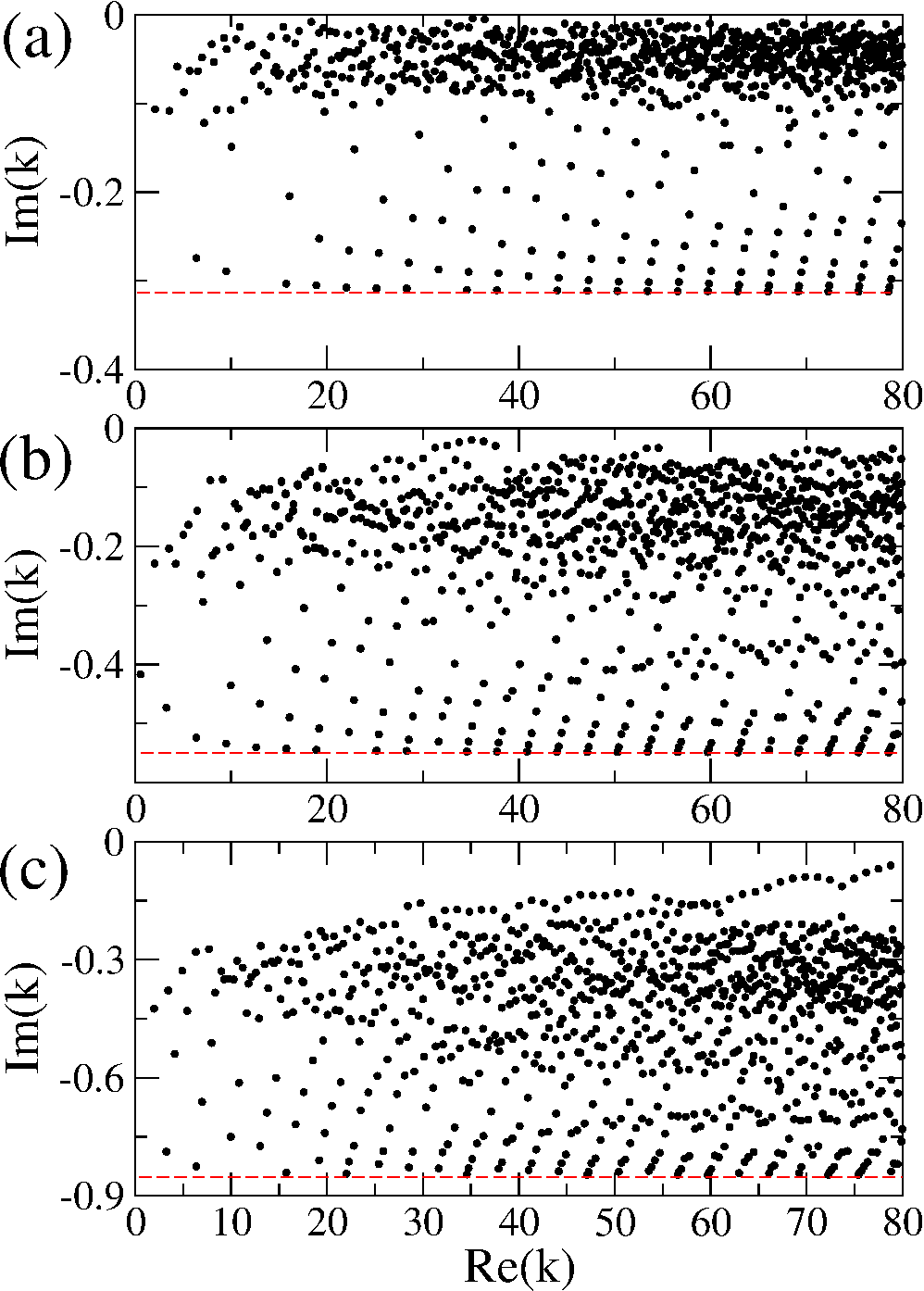}
\caption{(color online). \added{The complex eigenvalues of the modes of the stadium microcavities with (a) $n=3.3$, (b) $n=2.0$, and (c) $n=1.45$, for $\mathrm{Re}(k)<80$. Red dashed lines represent minimum imaginary values depending on the refractive indices.}}
\label{fig0}
\end{center}
\end{figure}

First, we consider the number of resonant modes in microcavities which can be inferred from the modified Weyl\'’s theorem \cite{Bal76, Lee04b}
\begin{equation}
N(k) \sim \frac{A k^2}{4 \pi} \mp \frac{L k}{4 \pi} + \cdots ,
\end{equation}
where $A$ and $L$ are the area and length of the boundary of the stadium microcavity, respectively, and the $-$($+$) refers to Dirichlet (Neumann) boundary conditions. If Dirichlet and Neumann boundary conditions are considered, the numbers of levels of which eigenmodes have even-even parities are given by $N(80) \sim 893$ and $N(80) \sim 926$, respectively, for $k < 80$. Since the boundary condition of a microcavity is that both wave function and its normal derivative on the boundary are not zero, the number of modes might be between those expected in the two boundary conditions. The numbers of the modes which have even-even parities in microcavities with $n=3.3$, $2.0$, and $1.45$ are 9$16$, $911$, and $918$, respectively. \added{Figure~\ref{fig0} show the complex eigenvalues of the modes of the stadium microcavities with three different refractive indices, $n=3.3$, $2.0$, and $1.45$, for $\mathrm{Re}(k)<80$. From the results, one can find the distributions of complex eigenvalues are very different depending on refractive indices.}

\begin{figure}
\begin{center}
\includegraphics[width=\figsizeone\textwidth]{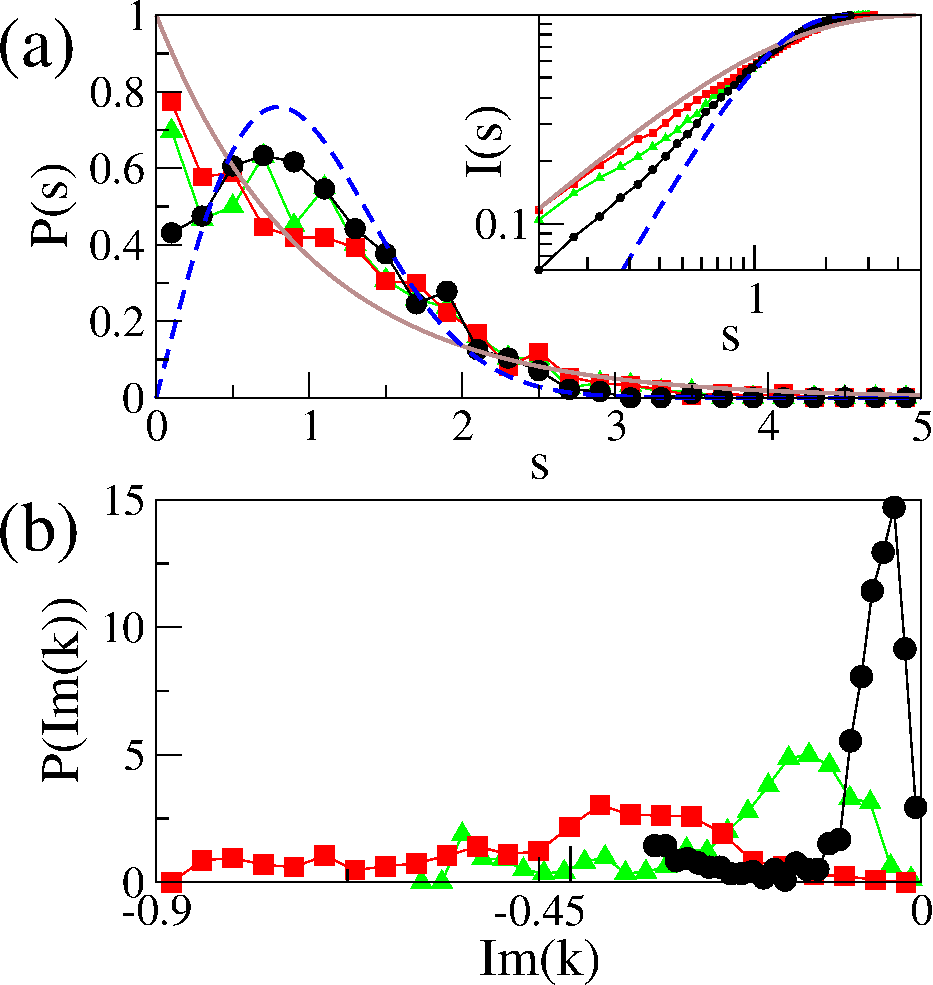}
\caption{(color online). (a) The level spacing distributions of real parts
and (b) the probability distributions of imaginary parts of eigenvalues for the stadium microcavities with $n=3.3$ (black circles), $2.0$ (green triangles), and $1.45$ (red rectangles), respectively.
The blue dashed and the brown curves represent the Wigner and the Poisson distributions, respectively.
(Inset) The integrated level spacing distributions of real parts of eigenvalues.}
\label{fig1}
\end{center}
\end{figure}

Next, the statistics of real parts of complex eigenvalues of the modes is considered. It is well known that the level spacing of a closed fully chaotic system exhibits the Wigner distribution.
Figure~\ref{fig1}(a) shows the level spacing distribution of the stadium microcavity with three different refractive indices, $n=3.3$, $2.0$, and $1.45$, for $\mathrm{Re}(k)<80$.
In dielectric microcavities the rays incident to the cavity boundary from the inside with the angle of incidence smaller
than the critical angle determined by $1/n$ can escape from the cavities so that the smaller $n$ the more rays leak out.
Thus decreasing $n$ makes the microcavities open wider.
The level spacing distribution of the stadium microcavity with large $n$, namely $n=3.3$
is similar to the Wigner distribution, which describes that of closed chaotic systems.
However, as the opening becomes larger ($n$ becomes smaller),
the level spacing distribution of $\mathrm{Re}(k)$'s changes from the Wigner ($n=3.3$) to Poisson ($n=1.45$) distribution.
For $n=2.0$ the level spacing distribution shows the intermediate behavior of $n=1.45$ and $3.3$.
The key nature of the Wigner distribution, a zero probability at $s=0$,
is ascribed to the interaction among energy levels giving rise to the avoided crossing.
In order to confirm the crossover behavior of level spacing distributions, we obtain the Brody parameter \cite{Bro73} for three level spacing distributnios in Fig.~\ref{fig1} (a). The Brody parameters are $0.552$, $0.225$, and $0.086$ in the cases of $n=3.3$, $2.0$, and $1.45$, respectively. As refractive indices decrease, the Brody parameters decrease. This confirms that there is crossover from Wigner to Poisson distribution as the refractive indices decrease.
As a result, the crossover from Wigner to Poisson distributions implies that the interaction between nearest neighboring modes are effectively reduced
as the refractive index decreases, i.e., opening strength increases.

The crossover can be understood from the viewpoint of short time dynamics of chaotic microcavities. In principle, the classical chaos is a long time (or infinite time) property and thus the ray dynamics in chaotic microcavities with low refractive indices is no longer chaotic because the rays inside chaotic microcavities leave the cavity after a few reflections. As a result, the short time dynamics of dielectric microcavities with low refractive indices can suppress the chaotic properties of the systems. However, this is not always true. For instance, the absorption loss of microcavities cannot suppress the chaotic properties of the systems in spite of the short time dynamics because the open properties are totally independent of the ray dynamics of the microcavities. We will discuss this more in the next section.

The distributions of the imaginary parts of the eigenvalues change from the narrow one with a small average value of
$\left|\mathrm{Im}(k)\right|$ ($n=3.3$) to the wider one with larger average ($n=1.45$) as shown in Fig.~\ref{fig1}(b). This is not surprising since the system becomes more lossy when the refractive index decreases. Similar behavior has been also observed in Ref. \cite{Wie08}. \added{The imaginary parts also have minimum values as functions of refractive indices for TM polarization cases as shown in Fig.~\ref{fig0} \cite{Ryu08}.}

\section{Matrix model}

Although we ascertain prominent changes of the statistical properties of eigenvalues of chaotic dielectric microcavities, it is difficult to prove directly
why the level spacing distritubtion changes from Wigner to Poisson distributions as opening strength increases, i.e., refractive index decreases in chaotic dielectric microcavities.
In closed chaotic billiard, RMT is useful for understanding statistical properties such as level spacing distributions.
In this section, we introduce a matrix model with randomly distributed elements corresponding to chaotic dielectric microcavities
and numerically study the statistical properties of eigenvalues of the matrix model.

An open quantum system has been often described by the effective non-Hermitian Hamiltonian \cite{Mah69, Miz93, Fyo97, Rot09}
\begin{equation}
H=H_{0}-i \gamma H_{1},
\label{eq1}
\end{equation}
where $H_{0}$ is the Hermitian Hamiltonian describing a closed quantum system.
The Hermitian $H_{1}$ has a specific algebraic structure of $AA^\dagger$, with $A$ being a $N \times M$ matrix of coupling amplitudes between
$N$ internal states and $M$ open channels.
The parameter $\gamma$ characterizes the strength of the interaction between the system and the environment.
The Hamiltonian (\ref{eq1}) has been studied in the limiting cases of small and large opening
in terms of the eigenstates of $H_0$ and $H_1$, respectively \cite{Rot09,Sto99,Sok89,Sok92}.
\added{In the case of small-rank of the channel space $M \ll N$ a very detailed characterization of its complex eigenvalues was obtained in the works of Fyodorov and collaborators \cite{Fyo96,Fyo99,Som99}, whereas the case of the comparable ranks of $H_0$ and $H_1$ when $M \sim N$ was analyzed in detail in Ref.\cite{Haa92} and Ref.\cite{Leh95}. A recent review of some works in that direction can be found in Ref.\cite{Fyo16}.}
In this work we use the eigenstates of $H_1$ as bases, which is meaningful if the opening is large enough so that $H_1$ dominates.
Diagonalizing $H_{1}$, Eq.~(\ref{eq1}) is rewritten as
\begin{eqnarray}
\label{eq2}
&H^{'}=H_0^{'}-i \gamma H_1^{'} = \\\nonumber
&
\left(
\begin{array}{cccc}
\epsilon_1 & c_{12} & \cdots & c_{1N} \\
c_{12}^{*} & \epsilon_2 & \cdots & c_{2N} \\
\vdots & \vdots & \ddots & \vdots \\
c_{1N}^{*} & c_{2N}^{*} & \cdots & \epsilon_N \\
\end{array}
\right)
-i \gamma
\left(
\begin{array}{cccc}
\Gamma_1^{'} & 0 & \cdots & 0 \\
0 & \Gamma_2^{'} & \cdots & 0 \\
\vdots & \vdots & \ddots & \vdots \\
0 & 0 & \cdots & \Gamma_N^{'} \\
\end{array}
\right).
\end{eqnarray}

If the $H_0$ describes chaotic internal dynamics, it is represented by the RMT.
Thus $H'_0$ can also be represented by RMT in general the unitary transformation of certain Hermitian matrices described by the RMT results in those described by the RMT again;
$H'_0=UH_0U^+ \in R$  if $H_0 \in R$, where $R$ denotes the set of the matrices described by the RMT and $UH_1U^+=H'_1$.
It is noted that the ensemble class of $H'_0$ is not important in our works since opening is large enough so that $H'_1$ dominates, not $H'_0$.

\begin{figure}
\begin{center}
\includegraphics[width=\figsizetwo\textwidth]{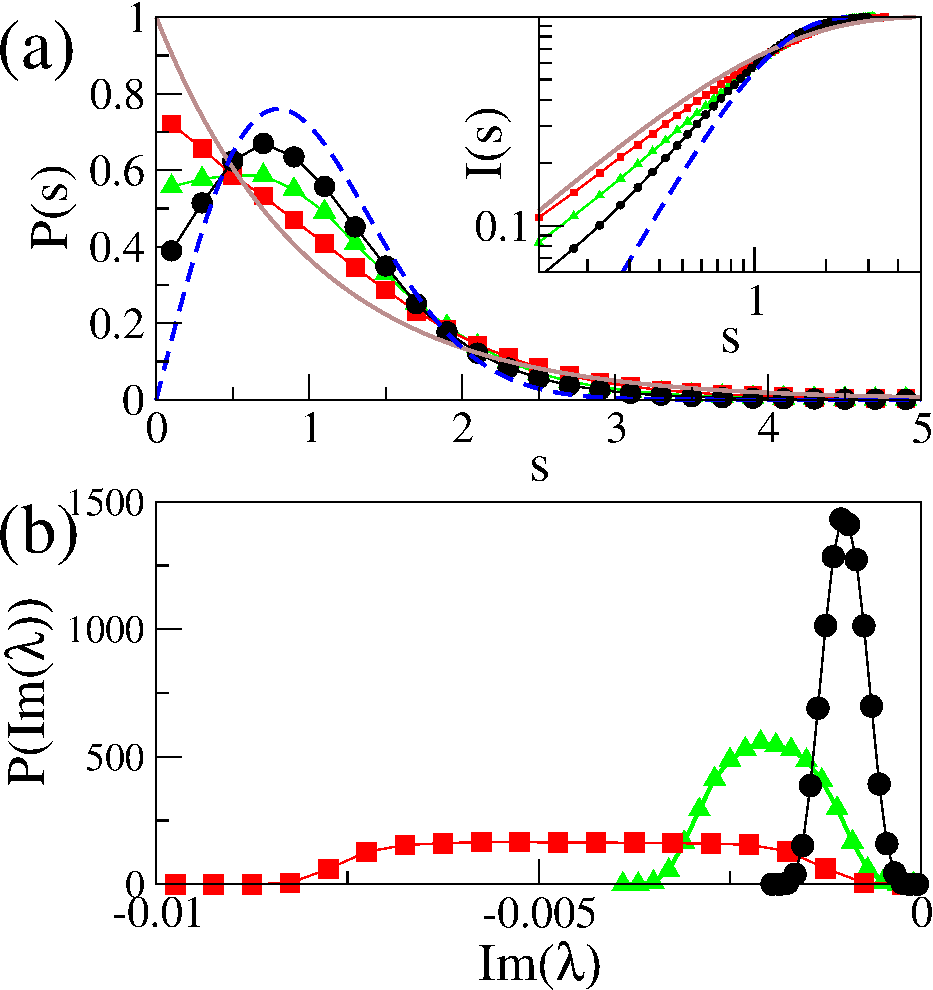}
\caption{(color online). (a) The level spacing distributions of real parts
and (b) the probability distributions of imaginary parts of eigenvalues of the $N \times N$ matrix model ($N=3000$) with $\gamma=0.002$ (black circles), $0.004$ (green triangles), and $0.009$ (red rectangles), respectively, with $c=0.001$.
The blue dashed and the brown curves represent the Wigner and the Poisson distributions, respectively.
(Inset) The integrated level spacing distributions of real parts of eigenvalues.
}
\label{fig2}
\end{center}
\end{figure}

The effective non-Hermitian Hamiltonian of Eq.~(\ref{eq1}) has been well known and widely studied.
In this work, the structure of the diagonal matrix $H'_1$ is the most important issue
because the structure of $H'_1$ plays a crucial role in determining the crossover behaviors.
Recall that here we construct a matrix model to describe the chaotic dielectric microcavities.
First we set $M=N$.
Strictly speaking microcavities have infinitely many scattering channels since the outside of the microcavities
is simply the continuum.
Thus we practically assume the number of the scattering channels is equal to that of internal states, namely $M=N$.
We also assume that $\Gamma'_m$ linearly increases with $m$, which is equivalent to the set of random numbers uniformly distributed within a certain range, namely $[0,1]$.
In fact, the reflectivity of a circular microcavity directly associated with loss gradually increases as $\sin\chi$ increases \cite{Hen02},
where $\chi$ denotes the angle of incidence of light from the inside at the boundary.
It is not exactly linear but at least monotonically increases in the bases labeled by $\sin\chi$,
which is directly related to  the azimuthal mode number {\it m} in a circular cavity.
Therefore, $H'$ is obtained by choosing the random numbers satisfying $-1 \leq \epsilon_j \leq 1$, $-c \leq c_{jk} \leq c$, and $0 \leq \Gamma_j \leq \gamma$, which are normalized by $\epsilon_j$.
The validity of such a structure of $H'$ is partly justified by the fact that the distribution of $\mathrm{Im}(\lambda)$ of the microcavity
is qualitatively similar to that of $H'$ for two cases, namely small and large opening,
except for the asymmetry of the distribution due to that of dielectric opening in the phase space, as shown in Fig.~\ref{fig1}(b) and Fig.~\ref{fig2}(b).

\added{It should be noted that if the number of open channels $M$ are smaller than the number of internal states $N$ for small coupling strength all eigenvalues are concentrated in a single cloud, but with increasing coupling strength the cloud of eigenvalues separates into two parts. The resonances corresponding to the fraction of coupled channels, i.e., $M/N$, have large negative imaginary parts, whereas the remaining resonances stay close to the real axis \cite{Haa92, Leh95}. If $M=N$, however, the eigenvalues form a single cloud corresponding to the probability distributions of imaginary parts of eigenvalues in Fig.~\ref{fig2}(b), irrespective of the value of coupling strength \cite{Haa91b}.}

The level spacing distribution of real parts of eigenvalues obtained from Eq.~(\ref{eq2}) with $c=0$ and $\gamma = 0$ exhibits the Poisson, which corresponds to the integrable system,
because of randomly distributed diagonal elements $\epsilon_j$ of $H^{'}_0$ \cite{Ber77b}.
As far as the closed system ($\gamma=0$) is concerned, it is well-known that the level spacing distribution evolves from Poisson to Wigner distributions as $c$ increases from zero.
The level spacing distribution when $c=0.001$ and $\gamma=0.0$ exhibits the Wigner, which means that the system is sufficiently chaotic.
Hereafter we fix $c=0.001$ for chaotic properties and vary $\gamma$ so as to control the opening strength.
It is shown in Fig.~\ref{fig2}(a) that with fixed c=0.001 the level spacing distribution of the real parts of eigenvalues evolves
from the Wigner ($\gamma=0.002$) to the Poisson ($\gamma=0.009$) as $\gamma$ increases.
Roughly speaking, the increase of $\gamma$ with fixed $c$ makes the diagonal part of the Hamiltonian (\ref{eq2}) dominate,
which effectively reduces the couplings (off-diagonal parts) among modes, compared with the diagonal parts.
The effective reduction of the couplings decreases the gaps of avoided level crossing among the neighboring eigenvalues
and thus causes the level spacing distribution to change from the Wigner to the Poisson.
\added{This result also coincides well with the statistical properties of the Ginibre ensemble which the statistics of complex eigenvalues tends to become when the number of open channels are sufficiently large \cite{Gin65, Gro88}. Ginibre eigenvalues repel each other only in the complex plane, whereas their projection on the real axis can be completely uncorrelated and do not show level repulsion at all.}

We emphasize that the change of the level spacing distribution is independent of the statistical properties of $H'_1$ unlike the Hermitian case.
In order to compare the non-Hermitian case of Eq.~(\ref{eq2}) with the Hermitian case,
let us consider the Hermitian Hamiltonian, $H=H_0 + \gamma H_1$, where the level spacing distribution of $H_0$ is Wigner.
In this case, if $\gamma$ is small, the eigenstates of $H$ are those of $H_0$ and the level spacing distribution of $H$ is Wigner.
As $\gamma$ increases, the eigenstates of $H$ become those of $H_1$.
If the level spacing distribution of $H_1$ is Wigner, the level spacing distribution of $H$ is Wigner even if $\gamma$ increases.
However, if the level spacing distribution of $H_1$ is Poisson, the level spacing distribution of $H$ is Poisson when $\gamma$ is sufficiently large.
Consequently, the level spacing distribution of $H$ is determined by that of $H_1$ if $\gamma$ is sufficiently large \cite{Len90, Len91}.
In open quantum systems with classically chaotic dynamics of Eq.~(\ref{eq2}), however, as $\gamma$ increases, the Wigner distributions always change into Poisson distributions,
regardless of whether the level spacing distribution of $H'_1$ is Wigner or Poisson because the level statistical distribution of $H'$ does not become that of $H'_1$ even if $\gamma$ is sufficiently large
but is determined by the randomly distributed diagonal elements $\epsilon_j$ of $H'_0$.

Large enough opening induces the change of localization properties of eigenstates of $H^{'}$ as well as statistical properites of eigenvalues (see Appendix B).
The previous experiments and numerical results in dielectric microcavities strongly support the eigenstates are often localized on the unstable periodic orbits or the unstable manifolds (see Appendix A).
Our matrix model, however, can not explain why it is so because of the lack of specific chaotic dynamics of dielectric microcavities.

It is noted that there are two kinds of loss for microcavity lasers, like a microwave cavity \cite{Sav06}.
One is the cavity loss which is due to refractive or tunneling emissions on the cavity boundary.
The second term of Eq.~(\ref{eq2}) was considered as the simple matrix model for the cavity loss.
The other is the absorption loss caused by the interaction with a material medium.
For the absorption loss, we have to consider a different matrix model from the $H_1^{'}$ of Eq.~(\ref{eq2}).
Considering only absorption loss without cavity loss, the decay rates of all basis states are same independent of individual properties of the basis states.
The additional term for absorption loss is $i \gamma_\alpha \Gamma_\alpha I$, where $\gamma_\alpha$ is coefficient of absorption loss
and $\gamma_\alpha \Gamma_\alpha$ is the constant decay rate of the all basis states.
The eigenvalues of $H^{'}=H_0^{'}-i \gamma_\alpha \Gamma_\alpha I$ equals to $\lambda - i \gamma_\alpha \Gamma_\alpha$ where $\lambda$ is the eigenvalues of $H_0^{'}$.
As $\gamma_\alpha$ increases, the real parts of eigenvalues of $H^{'}$ do not change and there is only lateral shift of probability distributions
of the imaginary parts without the change of shape of the distributions.
In addition, the part of identity matrix can not affect the eigenvectors and therefore,
as the $\gamma_\alpha$ increases, eigenvectors do not change in this model.
As a result, two kinds of loss in microcavity lasers, cavity loss and absorption loss,
play a very different role in the opening induced crossover behaviors.
This difference are originated from the fact that the crossover behaviors comes from increasing of not individual value $\Gamma_j$ but relative value $\Delta \Gamma_j$ (see Appendix C).

Finally, considering the dependence of the level spacing distributions on sizes of microcavities and matrix models,
there is no qualitative change of crossover behavior but the critical opening strength for the crossover will decrease as the sizes increases.
That is, if the sizes are larger, the level spacing distributions will be closer to the Poisson distributions when systems have same opening strengths,
$1/n$ in microcavities and $\gamma$ in matrix models.
Whether the level spacing distribution are close to the Wigner or Poisson distribution is determined by the ratio of the mean level spacing in $H_0^{'}$
to the mean difference between decay rates in $H_1^{'}$.
As sizes of microcavities and matrix models increase, the opening strength increases effectively
because the mean level spacing in $H_0^{'}$ decreases but the mean differnece between decay rates in $H_1^{'}$ does not change.

\section{Summary}

We have studied the variation of the statistics of eigenvalues in the Bunimovich stadium-shaped microcavities with different refractive indices.
The level spacing distributions change from Wigner to Poisson distributions and the probability distributions of decay rates become wider
as the refractive index of a microcavity decreases.
We have also proposed a non-hermitian matrix model with random elements, corresponding to the chaotic microcavity.
It provides us with plausible explanation on why the statistics of eigenvalues are changed according to the openness of the microcavity.

\section*{Acknowledgments}
We thank M. Choi, I. Kim, D. Lippolis, and S.-Y. Lee for discussions.
This work was supported by IBS-R024-D1. This work was supported by the National Research Foundation of Korea (NRF) grant (No.2016R1A2B4015978) and by a grant from KyungHee University in 2018 (KHU-20182175).

\appendix
\section{Localization of eigenmodes in a stadium microcavity}

\begin{figure*}
\begin{center}
\includegraphics[width=\figsizefive\textwidth]{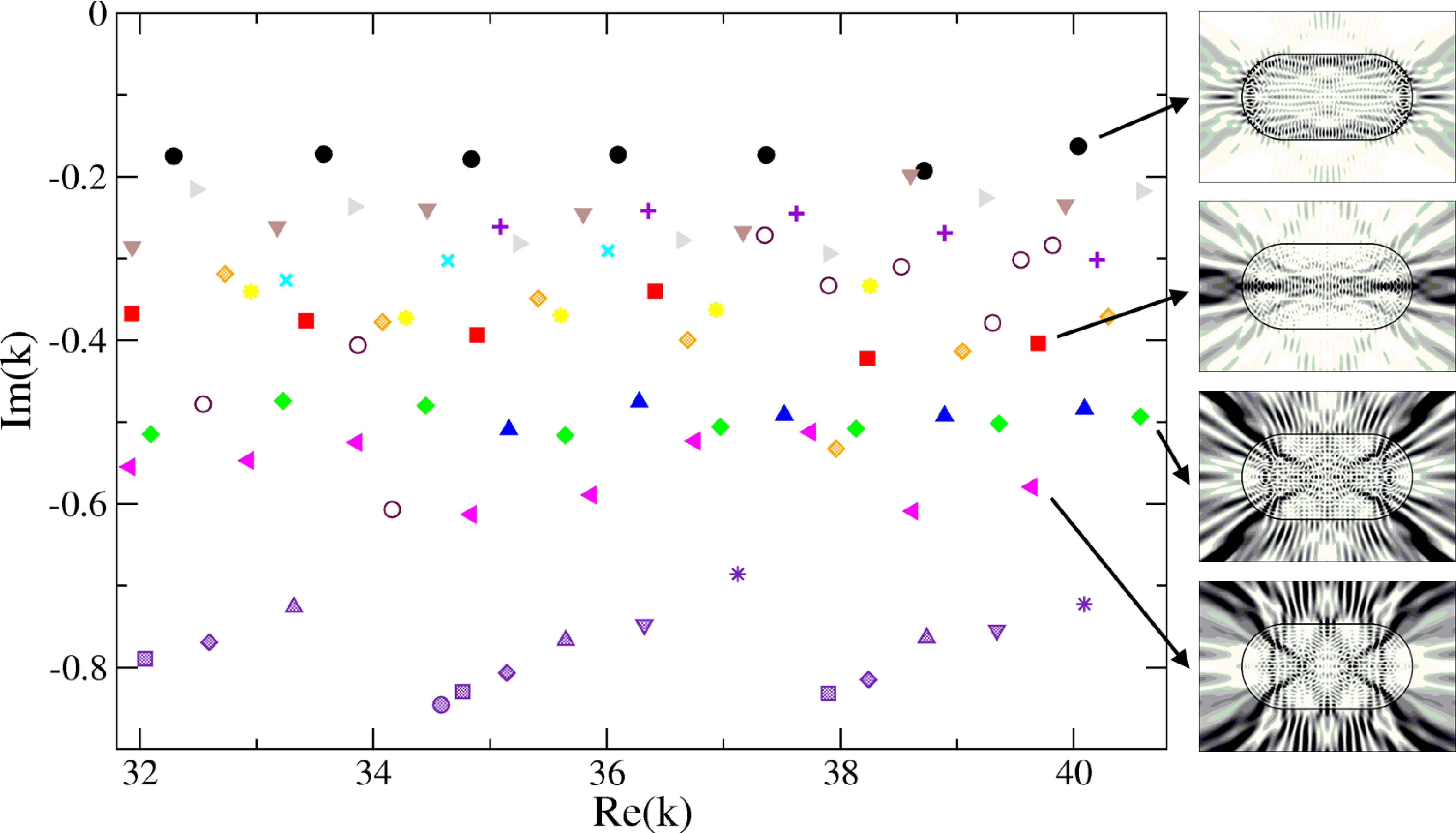}
\caption{(color online). The complex eigenvalues of the modes of the stadium microcavity with $n=1.45$.
The modes are grouped by the periodic orbits supporting their intensity pattern in coordinate space; the rectangle (black dots),
the arrowhead (brown triangle down and grey triangle right), the diamond (violet plus and cyan x), the triangle (yellow star),
the horizontal bouncing ball (red square), the fish (orange shaded diamond), the bowtie (blue triangle up and green diamond),
and the candy (magenta triangle left) shaped periodic orbits.
The several dots with indigo colors in low Q-region ($\mathrm{Im}(k)<-0.65$) represent the localized modes supported by the bouncing ball type orbits.
The maroon open dots represent delocalized or unclassified modes.
Four typical localized modes are presented in the right column in grayscale;
the rectangular, the horizontal bouncing ball, the bowtie, and the candy shaped orbits from the top.}
\label{fig5}
\end{center}
\end{figure*}

It is not easy to directly quantify the localization of statistically meaningful number of modes of the microcavities since it requires enormous numerical effort.
Instead, we show that for a given range of eigenvalues almost every modes are localized on the short periodic orbits.
In chaotic microcavities localization occurs in phase space.
More precisely eigenstates are localized on a certain periodic orbit, which consists of a group of modes with equidistant spacing in spectra,
as clearly shown in Fig.~\ref{fig5}, like the modes separated by a free spectral range in the Fabry-Perot cavity.
Thus the equidistant spacing itself directly implies
that the corresponding eigenmodes are localized in a certain periodic orbit with the well-defined path length $4\pi/\left<\Delta k\right>$, 
where $\Delta k$ is the spacing between the real parts of eigenvalues of two successive modes.

Figure~\ref{fig5} shows several sets of localized modes grouped by, if any, the corresponding periodic orbits.
The highest-Q modes (black dots), for instance, are localized on the rectangular periodic orbit.
Starting from the left ($k=32.289-i0.175$) the number of nodes of the spatial wavefunctions of these modes
increase from $22$ to $28$ in the quarter-stadium, from which one obtains the average $\left<\Delta k\right>$ is $1.292$.
This is well fitted with $1.301$ expected from quantization of the path length $l$ of the rectangular periodic orbit,
$\Delta k^{*}=4\pi/l$.
One finds that the spacing $\Delta k$ is almost equidistant
since the difference $\alpha = \left|(\left<\Delta k\right> - \Delta k^{*})/\Delta k^{*}\right|$ is $0.007$
and the standard deviation of $\Delta k$, denoted as $\sigma$, is $0.038$ (see Table~\ref{t1}).
It implies that the modes are strongly localized on the rectangular periodic orbit.

Besides the highest-Q modes, most of the modes in Fig.~\ref{fig5} are also localized on the short periodic orbits as summarized in Table~\ref{t1}.
Note that both $\alpha$ and $\sigma$ of all the modes are small enough to prove the equidistant spacing.
The group D has a rather larger $\alpha$ with small $\sigma$, where the intensity of the wavefunction appears to be slightly deviated
from the corresponding exact periodic orbit, namely a diamond shape.
In fact, the diamond periodic orbit is located just near the critical angle,
in which the so-called quasi-scar \cite{Lee04} can play an important role to induce such a deviation. 
For low-Q ($\mathrm{Im}(k)<-0.65$), the modes are still found to be strongly localized on the so-called bouncing ball trajectories
which form marginally stable orbits to be separated from the other parts of phase space.
For small opening ($n=3.3$), we hardly find groups of modes with equidistant spacing (not shown here)
so as to mostly observe chaotic-like states rather than localized ones which strongly localized on one periodic orbit.
For $n=2.0$ the intermediate behavior of $n=1.45$ and $3.3$ takes place (not shown here).

\begin{table*}
\caption{$\left<\Delta k\right>$, $\Delta k^{*}$, $\alpha$, and $\sigma$ of each group of the modes shown in Fig.~\ref{fig5} are presented with abbreviation; 
rectangle (R), arrowhead (A and A2), diamond (D and D2), triangle (T), horizontal bouncing ball (HBB), fish (F), bowtie (B and B2), and candy (C).}
\small
\begin{tabular}{ccccccccccccccccccccccc}
\hline
\hline
 & R & A & A2 & D & D2 & T & HBB & F & B & B2 & C \\
\hline
$\left<\Delta k\right>$ & 1.292 & 1.332 & 1.352 & 1.278 & 1.379 & 1.326 & 1.553 & 1.261 & 1.233 & 1.211 & 0.966 \\
$\Delta k^{*}$ & 1.301 & 1.360 & 1.360 & 1.405 & 1.405 & 1.400 & 1.571 & 1.364 & 1.209 & 1.209 & 0.952 \\
$\alpha$ & 0.007 & 0.021 & 0.006 & 0.090 & 0.019 & 0.053 & 0.011 & 0.076 & 0.020 & 0.002 & 0.015 \\
$\sigma$ & 0.038 & 0.067 & 0.054 & 0.018 & 0.013 & 0.005 & 0.152 & 0.095 & 0.107 & 0.061 & 0.060 \\
\hline
\hline
\end{tabular}
\label{t1}
\end{table*}

\section{Change of eigenstates of $H'$ in the matrix model}

\begin{figure}
\begin{center}
\includegraphics[width=\figsizethree\textwidth]{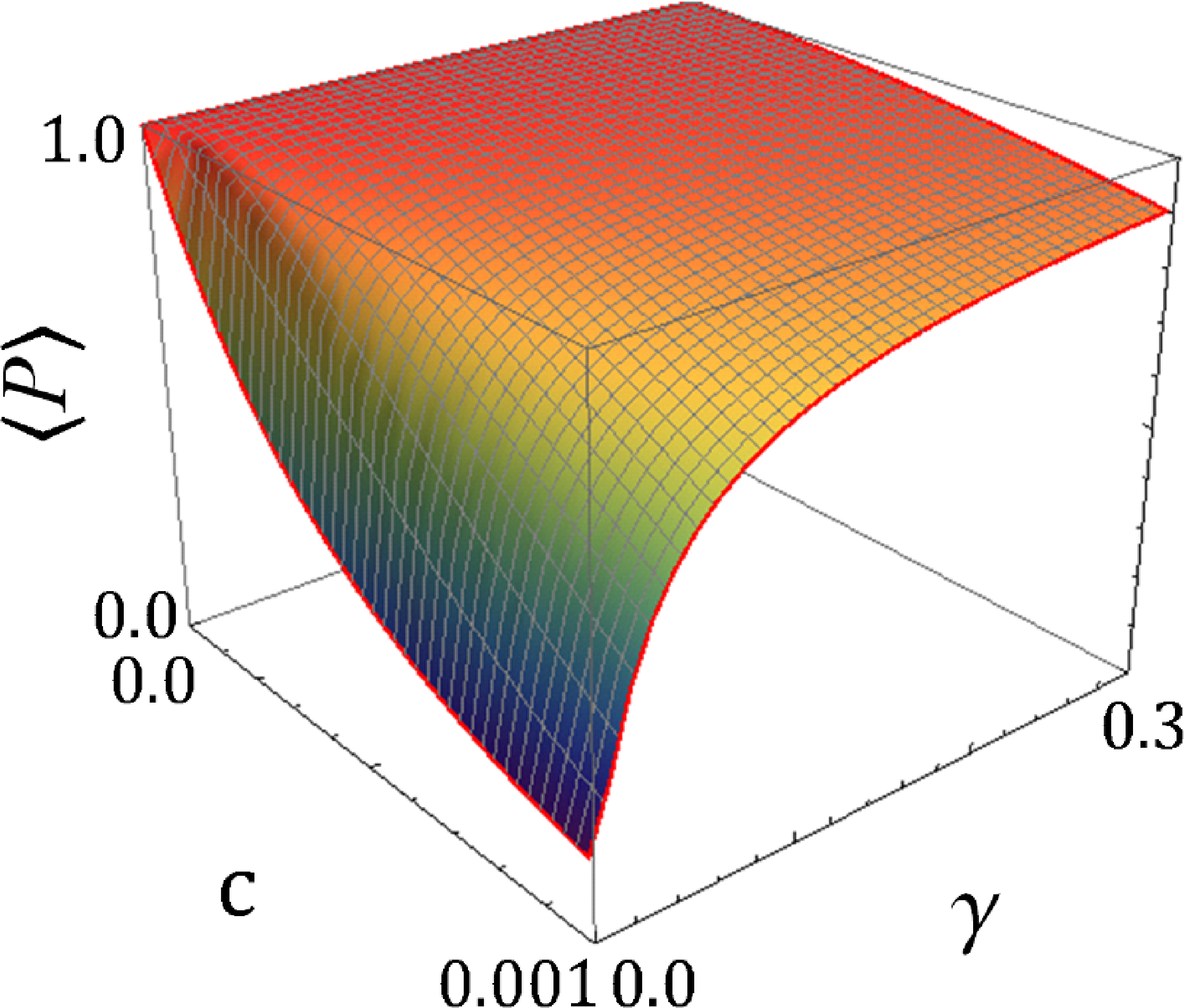}
\caption{(color online). The AIPR $\left<P\right>$ for the $N \times N$ Hamiltonian ($N=3000$) in terms of $c$ and $\gamma$.}
\label{fig3}
\end{center}
\end{figure}

We consider change of eigenstates of $H'$ in the matrix model.
In Eq.~(\ref{eq2}), it is obvious that if the opening is large enough so as to be $\Gamma_j \gg \left|c_{jk}\right|$ all the eigenstates of $H^{'}$ are
almost equivalent to those of $H^{'}_1$.
To measure how the eigenstates of $H'$ and $H'_1$ are identical, we introduce the average inverse participation ratio (AIPR),
defined as $\left<P\right>= \Sigma_i \left[ \frac{\Sigma_j |a^i_j|^{4}}{(\Sigma_j |a^i_j|^{2})^2} \right]/N$ ($1/N \le \left<P\right> \le 1$)
where $a^i_j$ is the $j$th element of the $i$th eigenstate \cite{Kap98a,Kap98b}.
The larger $\left<P\right>$, the more similar the eigenstates of $H'$ to those of $H'_1$
because eigenstates of $H'$ are localized on basis which are eigenstates of $H'_1$.
Figure~\ref{fig3} presents the AIPR in terms of $c$ and $\gamma$.
When the system is closed, i.e. $\gamma =0$, the AIPR monotonically decreases so that the eigenstates become mixed in terms of eigenstates of $H'_1$ as $c$ increases. 
However, if the system is open, for a given $c$ the AIPR monotonically increases to approach one
so that the eigenstates of $H'$ change from those of $H'_0$ to those of $H'_1$ as $\gamma$ increases as shown in Fig.~\ref{fig3}.
Most eigenstates of $H'$ become those of $H'_1$ if the opening ($\gamma$) is sufficiently larger than the coupling ($c$).
As a result, the change of statistical properties of eigenstates is accompany with that of the level statistical distribution of our matrix model.

\section{$2 \times 2$ matrix model}

Now let us consider the simplest case of Eq.~(\ref{eq2}) with $N=2$,
i.e., $2 \times 2$ matrix model.
It provides more precise criteria on the crossover behavior of eigenvalues and eigenstates in small and large opening cases.
This $2 \times 2$ matrix model has been widely used
for studying two interacting modes of open quantum systems to successfully explain various related experimental and theoretical phenomena
\cite{Dem01,Dem03,Wie06,Lee09}.

Assuming non-zero $c_{12}$ is real for simplicity, the eigenvalues are
$\lambda_{\pm}=\left(\epsilon_{1}+\epsilon_{2}-i\left(\Gamma_{1}+\Gamma_{2}\right)
\pm c \sqrt{\square}\right)/2$
and the corresponding non-normalized eigenstates are
\begin{equation}
\left(
\begin{array}{c}
\frac{1}{2}\left(-\frac{\Delta\epsilon}{c}+i\frac{\Delta\Gamma}{c} \pm\sqrt{\square}\right) \\
1 \\
\end{array}
\right),
\label{es1}
\end{equation}
where $\square=\left(\frac{\Delta\epsilon}{c}\right)^{2}-\left(\frac{\Delta\Gamma}{c}\right)^{2}+4-2i\frac{\Delta\epsilon\Delta\Gamma}{c^2}$.
We set $\Delta \epsilon=\epsilon_{2}-\epsilon_{1}$, $\Delta \Gamma=\Gamma_{2}-\Gamma_{1}$, and $c_{12}=c$, respectively.
First we consider the case of $\left|\frac{\Delta\Gamma}{c}\right| \ll 2$.
At $\Delta\epsilon=0$, the real parts of eigenvalues split into two different values
corresponding to the Wigner distribution of level spacing of $N \times N$ matrix model which shows a maximal peak around the mean level spacing.
The same imaginary parts are also related to the narrow distribution of imaginary parts of $N \times N$ matrix model.
The eigenstates of Eq.~(\ref{es1}) give rise to $\frac{1}{\sqrt{2}}(\pm 1, 1)^T$ representing the perfectly mixed states,
in the sense that it appears to be uniformly distributed over two bases, $(1,0)^T$ and $(0,1)^T$.
Next we consider the case of $\left|\frac{\Delta\Gamma}{c}\right| \gg 2$.
At $\Delta\epsilon=0$, the same real parts of eigenvalues correspond to the Poisson distribution of level spacing of $N \times N$ matrix model.
The two different imaginary parts are also related to the wide distribution of imaginary parts of $N \times N$ matrix model.
The eigenstates of Eq.~(\ref{es1}) are $(i,0)^T$ and $(0,1)^T$ representing the pure states.

\begin{figure}
\begin{center}
\includegraphics[width=\figsizethree\textwidth]{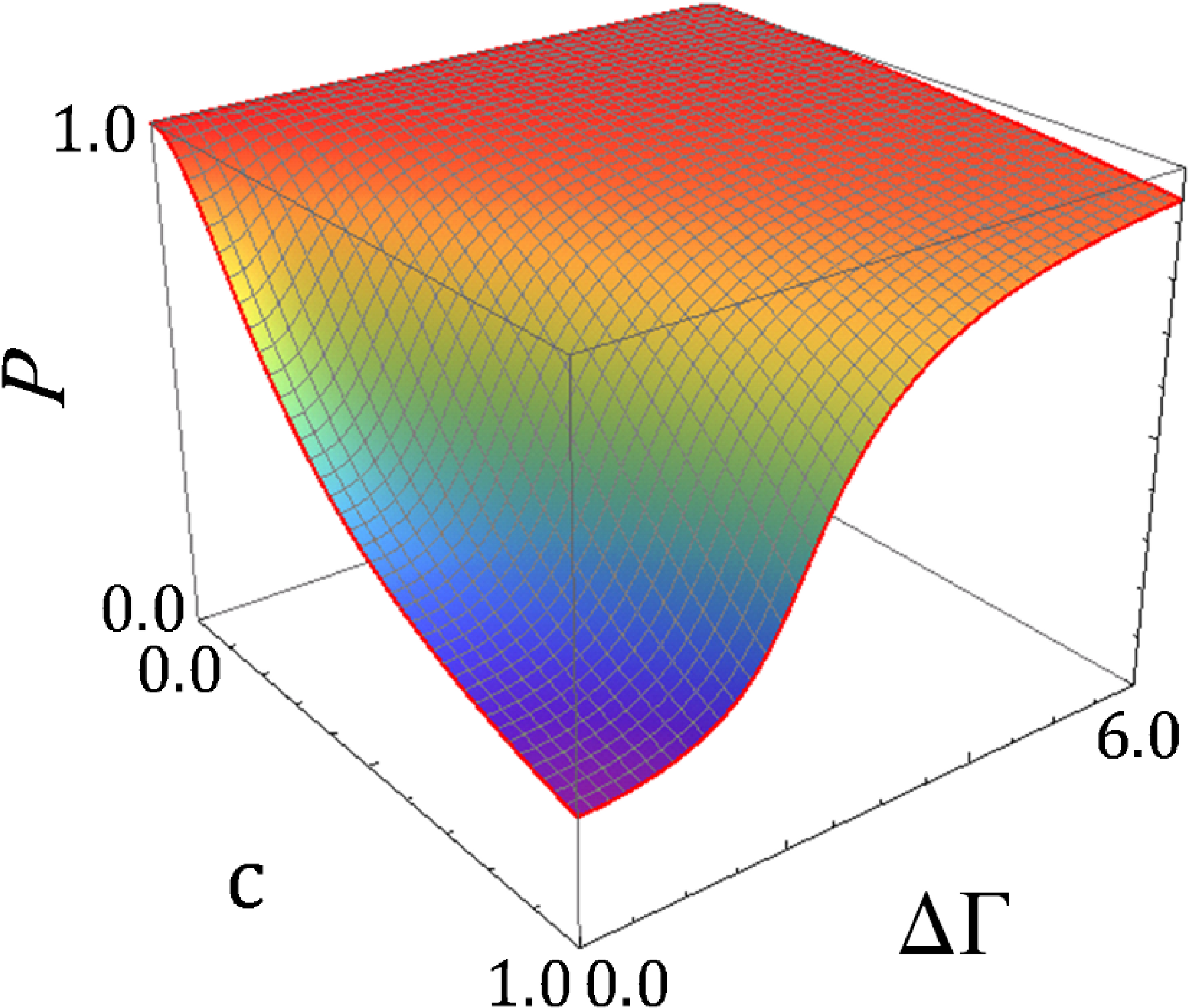}
\caption{(color online) The purity factor $P$ for the $2 \times 2$ matrix model in terms of $c$ and $\Delta \Gamma$.
}
\label{fig4}
\end{center}
\end{figure}

In order to quantitatively study the degree of change of the eigenstates, we define the purity factor $P$ as
$2R-1$ with $R=\mathrm{max}(|a_1|^2,|a_2|^2)$ when the normalized eigenstate is written as $(a_1,a_2)^T$.
The perfect pureness and the perfect mixing occur at $P=1$ and $P=0$, respectively.
Figure~\ref{fig4} shows the purity factor $P$ in terms of $c$ and $\Delta \Gamma$ when $\Delta \epsilon=0.5$.
If the system is closed, i.e., $\Gamma_1=\Gamma_2=0$, it is shown from Eq.~(\ref{es1}) that the eigenstates are pure
for $\left|c\right| \ll \left|\Delta\epsilon\right|$,
while mixed for $\left|c\right| \gg \left|\Delta\epsilon\right|$.
Once the system is opened, $\Delta\Gamma$ should be additionally taken into account;
Even if $\left|c\right|$ is large enough so as to be mixed in the closed system,
with $\left|\Delta\Gamma\right| \gg 2\left|c\right|$ the state appears to be pure.
In Fig.~\ref{fig4}, the purity factor $P$ in $2 \times 2$ matrix model increases monotonically with $\Delta \Gamma$ at any $c$,
which is qualitatively similar to the AIPR $\left<P\right>$ in $N \times N$ matrix model.
It leads us to conclude that any mixed state is transformed into a pure state as the opening increases
because generally speaking the larger the opening, the larger $\Delta \Gamma$.
However, it is emphasized that it is not $\Gamma_1$ and $\Gamma_2$ but the difference between them to avoid mixing.

\end{document}